
%
%
\input harvmac
%
%
%
%
\ifx\answ\bigans
\else
\output={
  \almostshipout{\leftline{\vbox{\pagebody\makefootline}}}\advancepageno
}
\fi
%
%
%

%
%
\def\doe{\#DOE-FG03-90ER40546}
\def\pyiam{PHY-8958081}

%
%
\def\UCSD#1#2{\noindent#1\hfill #2%
\bigskip\supereject\global\hsize=\hsbody%
\footline={\hss\tenrm\folio\hss}}
%
%
\def\abstract#1{\centerline{\bf Abstract}\nobreak\medskip\nobreak\par #1}
%
%
%
%
\edef\tfontsize{ scaled\magstep3}
 \tfontsize  \tfontsize
 \tfontsize \font\titlei=cmmi10 \tfontsize
\font\titleis=cmmi7 \tfontsize \font\titleiss=cmmi5 \tfontsize
\font\titlesy=cmsy10 \tfontsize \font\titlesys=cmsy7 \tfontsize
\font\titlesyss=cmsy5 \tfontsize  \tfontsize
\skewchar\titlei='177 \skewchar\titleis='177 \skewchar\titleiss='177
\skewchar\titlesy='60 \skewchar\titlesys='60 \skewchar\titlesyss='60
%
%
%
%
%
\def\inv{^{\raise.15ex\hbox{${\scriptscriptstyle -}$}\kern-.05em 1}}
\def\lbar{{\lower.35ex\hbox{$\mathchar'26$}\mkern-10mu\lambda}} 

%
%
%
%
\def\slash#1{\rlap{$#1$}/} 
\def\dsl{\,\raise.15ex\hbox{/}\mkern-13.5mu D} 
\def\delsl{\raise.15ex\hbox{/}\kern-.57em\partial}
\def\Ksl{\hbox{/\kern-.6000em\rm K}}
\def\Asl{\hbox{/\kern-.6500em \rm A}}
\def\Dsl{\hbox{/\kern-.6000em\rm D}} 
\def\Qsl{\hbox{/\kern-.6000em\rm Q}}
\def\gradsl{\hbox{/\kern-.6500em$\nabla$}}
%
%
\def\lspace{\ifx\answ\bigans{}\else\qquad\fi}
\def\lbspace{\ifx\answ\bigans{}\else\hskip-.2in\fi} 
%
%
\def\boxeqn#1{\vcenter{\vbox{\hrule\hbox{\vrule\kern3pt\vbox{\kern3pt
        \hbox{${\displaystyle #1}$}\kern3pt}\kern3pt\vrule}\hrule}}}
%
%
\def\mbox#1#2{\vcenter{\hrule \hbox{\vrule height#2in
\kern#1in \vrule} \hrule}}
%
%
%
%

   \def\CL{{\cal L}}
  \def\CO{{\cal O}}

%
%
%
%
%

%

\def\bar#1{\overline{#1}}
\def\vev#1{\left\langle #1 \right\rangle}
\def\bra#1{\left\langle #1\right|}
\def\ket#1{\left| #1\right\rangle}
\def\abs#1{\left| #1\right|}

\def\darr#1{\raise1.5ex\hbox{$\leftrightarrow$}\mkern-16.5mu #1}

%
%
\def\frac#1#2{{\textstyle{#1\over #2}}} 
%
%
%
%

\def\Tr{\mathop{\rm Tr}}

\def\MeV{{\rm MeV}}

%
%
%
%

%
%
\def\ltap{\ \raise.3ex\hbox{$<$\kern-.75em\lower1ex\hbox{$\sim$}}\ }
\def\gtap{\ \raise.3ex\hbox{$>$\kern-.75em\lower1ex\hbox{$\sim$}}\ }
\def\gl{\ \raise.5ex\hbox{$>$}\kern-.8em\lower.5ex\hbox{$<$}\ }
\def\roughly#1{\raise.3ex\hbox{$#1$\kern-.75em\lower1ex\hbox{$\sim$}}}
%
%
\def\ie{\hbox{\it i.e.}}        
\def\eg{\hbox{\it e.g.}}        
\def\etal{\hbox{\it et al.}}

\def\np#1#2#3{Nucl. Phys. B{#1} (#2) #3}
\def\pl#1#2#3{Phys. Lett. {#1}B (#2) #3}

\def\physrev#1#2#3{Phys. Rev. {#1} (#2) #3}

\relax

%
%
\noblackbox
\def\lc{\Lambda_c}
\def\lb{\Lambda_b}
\def\lqcd{\Lambda_{\rm QCD}}
\def\twolr{SU(2)_L \times SU(2)_R}
\def\larr#1{\raise1.5ex\hbox{$\leftarrow$}\mkern-16.5mu #1}
\def\vprime{\vec v^{\,\prime}}
\def\pprime{\vec p^{\,\prime}}

\centerline{{\titlefont{The Baryon Isgur-Wise Function}}}
\bigskip
\centerline{{\titlefont{in the Large $N_c$ Limit}}}
\bigskip\medskip
\centerline{Elizabeth Jenkins and Aneesh V.~Manohar}
\smallskip
\centerline{{\sl Department of Physics, University of California at San
Diego, La Jolla, CA 92093}}
\bigskip
\centerline{Mark B.~Wise}
\smallskip
\centerline{{\sl California Institute of Technology, Pasadena, CA 91125}}
\vfill
\abstract{ In the large $N_c$ limit,
the $\lb$ and $\lc$ can be treated as bound states of chiral
solitons
and mesons containing a heavy quark. We show that the
soliton and heavy
meson are bound in an attractive harmonic oscillator potential.
The Isgur-Wise function for $\lb\rightarrow\lc\, e^-\,
\bar\nu_e$ decay is computed in the large $N_c$ limit.
Corrections to the form factor which depend on
$m_N/ m_Q$ can be summed exactly ($m_N$ and $m_Q$
are the nucleon and heavy quark masses).
We find that this symmetry breaking correction at zero recoil
is only $1\%$.
}
\vfill
\UCSD{\vbox{\hbox{UCSD/PTH
92-27}\hbox{CALT-68-1809}\hbox{hep-ph/9208248}}}{August 1992}

\newsec{Introduction}
In the heavy quark limit, the form factors for semileptonic $\lb\rightarrow \lc
\,e^- \,\bar \nu_e$ decay \ref\iwg{N. Isgur and M.B. Wise,
\np{348}{1991}{278}\semi H. Georgi, \np{348}{1991}{293}\semi
F. Hussain, J.G. Korner, M. Kramer, and G. Thompson, Z. Phys. C51 (1991)
321.}\
are characterized by a single universal function
$\eta(v\cdot v')$
\eqn\explicit{
\bra{\lc(v',s')}\bar c\,\gamma^\mu\left(1-\gamma_5\right)b\ket{\lb(v,s)} =
\eta(v
\cdot v')\ \bar u(v',s')\gamma^\mu\left(1-\gamma_5\right) u(v,s),
}
where $v^\mu$ and $v'^{\mu}$ are the four-velocities of the
$\lb$ and $\lc$, respectively.
The Isgur-Wise function $\eta(v\cdot v')$ \ref\iw{N. Isgur and M.B.
Wise, \pl{232}{1989}{113}, \pl{237}{1990}{527}.}\
has logarithmic dependence on the
heavy $b$ and $c$ quark masses which is calculable using perturbative QCD
methods. The quark mass dependence can be put into a multiplicative
factor \ref\fggw{A.F. Falk, H. Georgi, B. Grinstein, and M.B. Wise,
\np{343}{1990}{1}.}
\eqn\scaling{
\eta(v\cdot v') = C_{cb}(v\cdot v')\ \eta_0(v\cdot v'),
}
where
\eqn\coef{
C_{cb}(v \cdot v') = \left[{\alpha_s(m_b)\over\alpha_s(m_c)}\right]^{-6/25}
\left[{\alpha_s(m_c)\over\alpha_s(\mu)}\right]^{a_L(v\cdot v')},
}
and
\eqn\al{
a_L(v\cdot v') = {8\over 27} \left[ v\cdot v'\ r(v\cdot v') -1 \right],
}
\eqn\rfn{
r(v\cdot v') = {1\over \sqrt{(v\cdot v')^2-1}}\ln\left(v\cdot v' +
\sqrt{(v\cdot v')^2-1}\right).
}
For very large heavy quark masses (and $\mu$ of order the QCD scale), $C_{cb}(v
\cdot v')$
has a rapid dependence on $v\cdot v'$.
The function $\eta_0(v\cdot v')$ is determined by low-momentum strong
interaction physics. It depends on the subtraction point $\mu$ in a way that
cancels the subtraction point dependence of $C_{cb}(v\cdot v')$. At zero
recoil, \ie\ $v\cdot v' =1$, $\eta_0$ is independent of $\mu$ and
is normalized to unity \iw \ref\nw{S. Nussinov and W. Wetzel,
\physrev{D36}{1987}{130}.}\ref\vs{M.B. Voloshin and M.A. Shifman, Sov.
J. Nucl. Phys. 47 (1988) 199.}\ by heavy quark flavor
symmetry,
\eqn\etanorm{
\eta_0(1) = 1.
}

Some of the low momentum properties of QCD are determined by its symmetries
(\eg\ chiral symmetry and heavy quark symmetry). Those nonperturbative
aspects of the theory which are not determined by symmetries
cannot be treated using perturbation theory in
the strong coupling constant.
QCD, however, does have an expansion parameter which can be used
to study low momentum features of the strong interactions analytically.
In the limit that the number of colors $N_c$ is
large, the theory simplifies and many predictions are possible
\ref\thooft{G. 't Hooft, \np{72}{1974}{461}, \np{74}{1974}{461}.}. The main
purpose of this paper is to examine $\eta_0(v\cdot v')$ in the large $N_c$
limit.

In the large $N_c$ limit, baryons containing light $u$ and $d$ quarks can be
viewed as solitons~\ref\skyrme{T.H.R. Skyrme, Proc. Roy. Soc. A260 (1961)
127.}\ref\witten{E. Witten, \np{223}{1983}{433}.}
of the nonlinear chiral Lagrangian for pion
self-interactions. Baryons containing a single heavy $c$ (or $b$) quark and
light $u$ and $d$ quarks are then described as bound states of these solitons
with $D$ and $D^*$ mesons (or $\bar B$ and $\bar B^*$ mesons)
\ref\ck{C.G. Callan and I. Klebanov, \np{262}{1985}{365}.}\ref\jmw{E.
Jenkins, A.V. Manohar and
M.B. Wise, Caltech Preprint CALT-68-1783 (1992).}.
The large $N_c$ behavior of $\eta_0(v \cdot v')$ can be determined
using the bound state wavefunctions of the $\Lambda_b$ and $\Lambda_c$.
For large $N_c$, the function $\eta_0(v \cdot v')$ is strongly peaked
about zero recoil since any velocity change must be transferred to
$\sim N_c$ light quarks.  Independent of the details of the bound
state approach, we find that
\eqn\neta{
\eta_0(v \cdot v') = {\rm exp}[- \lambda N_c^{3/2} (v \cdot v' - 1 ) ],
}
where $\lambda$ is a constant of order unity. This equation is valid for
$v\cdot v'-1$ of order $N_c^{-3/2}$. In this kinematic region, $\eta_0$
falls from unity to a very small quantity. The derivation of this
result is the main purpose of this paper. In addition, the effect on
$\eta_0$ of corrections to the heavy quark limit that depend on
$m_N/m_Q$ is examined.
Some features of the soliton picture of heavy baryons not discussed in previous
work on this subject will also
be derived here. In Refs.~\jmw\ref\glm{{Z. Guralnik,
M. Luke and A.V. Manohar, UCSD Preprint UCSD/PTH 92-24 (1992).}}
it was shown that the
leading term in the chiral Lagrangian for heavy-meson--pion interactions gives
rise to a heavy-meson--soliton potential that is attractive at the
origin in the $\Lambda_Q$, $\Sigma_Q$ and $\Sigma_Q^*$ channels. We
show in this paper that the curvature of the soliton--heavy-meson potential is
positive, indicating that the origin is a stable minimum of the potential
energy for these channels. The curvaturve is negative for the exotic
channels.

\newsec{$\Lambda_Q$ as a Heavy Meson-Soliton Bound State}
The starting point for discussing soliton--heavy-meson bound states is the
chiral Lagrangian for the interactions of mesons containing a heavy quark $Q$
with pions \ref\chiralrefs{M. Wise, \physrev{D45}{1992}{2118}\semi
G. Burdman and J.F. Donoghue, \pl{280}{1992}{287}\semi
T.M. Yan, H.Y. Cheng, C.Y. Cheung, G.L. Lin,
Y.C. Lin and H.L. Yu, \physrev{D46}{1992}{1148}.}.
In the limit $m_Q\rightarrow\infty$, the total angular momentum of
the light degrees of freedom, $\vec S_\ell$, is a
symmetry generator. The lowest
mass mesons with $Q\bar q_a$ ($q_1=u$, $q_2=d$) flavor quantum numbers have
$s_\ell=1/2$ and form a degenerate doublet consisting of pseudoscalar
and vector
mesons. In the case $Q=c$, these are the $D$ and $D^*$ mesons, and in the case
$Q=b$, these are the $\bar B$ and $\bar B^*$ mesons.

It is convenient to combine the fields $P_a$ and $P_{a\mu}^*$ for the ground
state $s_\ell=1/2$ mesons into the bispinor matrix
\eqn\hmatrix{
H_a = {(1+\slash v)\over 2}\left[ P^*_{a\mu}\gamma^\mu - P_a \gamma_5 \right],
}
where $v^\mu$ is the heavy quark four velocity, and $v^2=1$. The vector meson
field is constrained to satisfy $v^\mu P^*_{a\mu}=0$. In this section,
we work in
the rest frame of the heavy meson, $v^\mu=(1,\vec 0)$.
Under the heavy quark
spin symmetry,
\eqn\hspin{
H_a\rightarrow S H_a,
}
where $S\in SU(2)_v$ is the heavy quark spin transformation. The transformation
property of $H$ under $\twolr$ chiral symmetry has an arbitrariness associated
with field redefinitions. We will use the basis chosen in
Ref.~\jmw,\footnote{*}{The computations are repeated for the $\xi$ basis
in the appendix.} with the transformation rule
\eqn\hprime{
H_a\rightarrow (H R^\dagger)_a,
}
under $\twolr$, where $R\in
SU(2)_R$.
It is also convenient to introduce the field
\eqn\hdagger{
\bar H^a = \gamma^0 H_a^\dagger \gamma^0  =\left[ P^{*\dagger}_{a\mu}\gamma^\mu
+ P_a^\dagger \gamma_5 \right] {(1+\slash v)\over 2}.
}
The Goldstone bosons occur in the field
\eqn\sigdef{
\Sigma=\exp\left({2 i M\over
f}\right),
}
where
\eqn\Mdef{
M=\left[\matrix{\pi^0/\sqrt2&\pi^+\cr\pi^-&-\pi^0/\sqrt2\cr}\right],
}
and $f\approx132$~MeV is the pion decay constant.
Under $\twolr$
\eqn\sigtrans{
\Sigma \rightarrow L \Sigma R^\dagger,
}
with $L\in SU(2)_L$ and $R\in SU(2)_R$.
Under parity,
\eqn\sigpar{
\Sigma(x^0,\vec x)\rightarrow \Sigma^\dagger(x^0,-\vec x),
}
since $M(x^0,\vec x)\rightarrow -M(x^0,-\vec x)$. If $H_a$ transforms under
chiral symmetry as in Eq.~\hprime, then the parity transform of $H$ must
transform under chiral symmetry with a factor of $L^\dagger$. Consequently, in
the basis we are using, the action of parity on the $H$ field is somewhat
unusual,
\eqn\hparity{
H_a(x^0,\vec x) \rightarrow \gamma^0 H_b(x^0,-\vec x) \,\gamma^0
\,\Sigma^{\dagger
b}{}_a(x^0,-\vec x).
}

The chiral Lagrangian density for heavy meson-pion strong interactions is
\eqn\hlag{
\CL = -i \Tr \bar H v\cdot \partial H + {i\over 2} \Tr \bar H H v^\mu
\Sigma^\dagger \partial_\mu \Sigma + {ig\over 2}
\Tr \bar H H \gamma^\mu\gamma_5 \Sigma^\dagger \partial_\mu \Sigma + \ldots,
}
where the ellipsis denotes the contribution of terms containing more
derivatives or factors of $1/m_Q$. The coefficient of the second term in the
Lagrangian density Eq.~\hlag\ is fixed (relative to the first) by parity
invariance. The coupling $g$ determines the $D^*\rightarrow D\pi$ decay rate.
Present experimental information on the $D^*$ width and the $D^*\rightarrow
D\pi$ branching ratio implies that $g^2<0.4$~\ref\accmor{The ACCMOR
collaboration (S. Barlag \etal), \pl{278}{1992}{480}.}.
The constituent quark model
predicts that $g$ is positive.

The soliton solution of the $\twolr$ chiral Lagrangian
for baryons containing $u$ and $d$ quarks is
\eqn\sol{
\Sigma = A(t)\ \Sigma_0(\vec x) A^{-1}(t),
}
where
\eqn\skyrme{
\Sigma_0 = \exp\left(i F(r)\ \hat x\cdot \vec\tau\right),
}
and $r=\abs{\vec x}$. $A(t)$ contains the dependence on the collective
coordinates associated with rotations and isospin transformations of the
soliton
solution. For solitons with baryon number one, $F(0)=-\pi$ and $F(\infty)=0$.
The
detailed shape of $F(r)$ depends on the chiral Lagrangian for pion self
interactions including terms with more than two derivatives. We expect that
$\Sigma_0(\vec x)$ has a power series expansion in $\vec x$. Consequently, the
even powers of $r$ must vanish when $F(r)$ is expanded in a power series in
$r$, \eg\ $F''(0)=0$.  The chiral Lagrangian for pion self interactions is of
order $N_c$. However, the chiral Lagrangian for heavy-meson--pion interactions
is only of order one. Thus, to leading order in $N_c$ the shape of the
soliton $F(r)$ is not altered by the presence of the heavy meson.

In the large $N_c$ limit, baryons containing light $u$ and $d$ quarks are very
heavy and time derivatives on the $\Sigma$ field can be neglected.
Consequently,
it is the interaction Hamiltonian
\eqn\hint{
H_I = -{ig\over 2}\int d^3 \vec x\ \Tr \bar H H
\gamma^j\gamma_5\Sigma^\dagger\partial_j
\Sigma + \ldots,
}
with $\Sigma$ given by Eqs.~\sol\ and \skyrme\
that determines the potential energy of a
configuration with a heavy meson at the origin and a baryon at position
$\vec x$. Neglecting operators with more than one derivative (the ellipsis in
Eq.~\hint), and expanding the interaction potential operator in $\vec x$ gives
\eqn\hexpand{\eqalign{
\hat V_I(\vec x) &= g\ S^j_{\ell H}\ I^k_{H}\
 \Tr A \tau^i A^{-1} \tau^k \left\{
\delta^{ij}\left[F'(0) - \frac23 r^2 \left[F'(0)\right]^3
+\frac16 r^2 F'''(0)\right]\right.\cr&\qquad\left. + x^i x^j\left[
\frac23 \left[F'(0)\right]^3 + \frac13 F'''(0)\right] +
\epsilon^{ijm} x^m \left[F'(0)\right]^2\right\}+\CO(x^3),
}}
where $S^j_{\ell H}$ denotes the angular momentum of the light degrees of
freedom of the heavy meson, and $I^k_H$ denotes the isospin of the heavy meson.
The interaction potential has terms which superficially have the wrong
parity, {\it e.g.} the term involving the $\epsilon$ symbol.
However, these terms are required because of the $\Sigma^\dagger$ factor in the
parity transformation of $H$ in Eq.~\hparity.

The $\Lambda_Q$ baryon
has isospin zero and total angular momentum of the light
degrees of freedom equal to zero. In the large $N_c$ limit, it arises from a
bound state of nucleons with $P$ and $P^*$ mesons. Baryons with $I>1/2$ such as
the $\Delta$ cannot produce a heavy baryon bound state with $I=0$. On
nucleon states, $\Tr A \tau^i A^{-1} \tau^k$ is equal to $-8 S^i_N I^k_N/3$
where $S^i_N$ is the spin of the nucleon, and $I^k_N$ is the isospin of the
nucleon \ref\anw{G.S. Adkins, C.R. Nappi, and E. Witten,
\np{228}{1983}{552}.}. Using this simplification, the potential operator
becomes
\eqn\potop{
\hat V_I(\vec x) = \hat V_I^{(0)} + \hat V_I^{(1)}+\hat V_I^{(2)}+\CO(x^3),
}
where $\hat V_I^{(n)}$ denotes the term of order $r^n$ in the potential
\eqn\potans{\eqalign{
\hat V_I^{(0)} &= -{\frac 8 3}\ g F'(0)\ \vec I_H\ \cdot \vec I_N
\ \vec S_{\ell H}\cdot \vec S_N,\cr
\noalign{\smallskip}
\hat V_I^{(1)} &= {\frac 8 3}\ g \left[F'(0)\right]^2\ \vec I_H \cdot \vec I_N
\ \vec x \cdot (\vec S_{\ell
H} \times \vec S_N),\cr
\noalign{\smallskip}
\hat V_I^{(2)} &= -{\frac 8 3}\ g\ \vec I_H \cdot \vec I_N \left\{ \vec
S_{\ell
H}\cdot \vec S_{N} \left[
-{\frac 2 3} r^2 \left[F'(0)\right]^3 + {\frac 1 6}r^2 F'''(0)\right]\right.\cr
&\left.\qquad\qquad+\ (\vec S_{\ell H}\cdot \vec x) (\vec S_N\cdot \vec
x)\left[
{\frac 2 3}
\left[F'(0)\right]^3 + {\frac 1 3} F'''(0)\right]\right\}.
}}
The potential $\hat V_I(\vec x)$ commutes with the total angular momentum of
the light degrees of freedom, $\vec L + \vec S_{\ell H} + \vec S_N$,
where $\vec L$ is the orbital
angular momentum.

To find the  ${\Lambda_Q}$ wavefunction $\Psi_\Lambda(\vec x)$
and its potential energy
$V_\Lambda(\vec x)$, the
potential energy operator must be diagonalized at each point $\vec x$ on the
product space of nucleon heavy meson states (\eg\
$\ket{p,\,\uparrow}\ket{P^*,\,-}$). It
is convenient to consider linear combinations of these product states that have
definite isospin $\vec I = \vec I_H + \vec I_N$, spin of the light degrees of
freedom, $\vec S_\ell =\vec S_{\ell H} + \vec S_N$,
and total spin $\vec S = \vec
S_Q + \vec S_\ell$. These states are labeled $\ket{I,s,s_\ell}$. It is
straightforward to diagonalize $\hat V_I(\vec x)$ in this basis, and we find
that
\eqn\lamstate{
\Psi_{\Lambda}(\vec x) = \left[ 1 -  F'(0)\ \vec x\cdot(\vec S_{\ell
H}
\times \vec
S_N) \right]
\ket{0,\frac 1 2,0}\phi(\vec x),
}
is an eigenstate of $\hat V_I(\vec x)$ with eigenvalue
\eqn\pot{\eqalign{
V_\Lambda(\vec x) &= -\frac 3 2 g F'(0) + g r^2 \left[\frac 1 6
\left[F'(0)\right]^3 - \frac 5 {12} F'''(0)\right]\cr
&= -\frac 3 2 g F'(0) + \frac 1 2 \kappa r^2,
}}
where $\kappa$ is defined by
\eqn\kapdef{
\kappa = g \left[ \frac 1 3 \left[F'(0)\right]^3 -
\frac 5 6 F'''(0) \right].
}
The form of the wavefunction in Eq.~\lamstate\ can also be found
using Eq.~\hparity\ and demanding that it has definite parity.
The factor in square brackets in Eq.~\lamstate\ compensates for the factor of
$\Sigma^\dagger$ in Eq.~\hparity.
In Eqs.~\lamstate\ and \pot, $r\lqcd$ is treated as a small
quantity\footnote{*}{$\lqcd$ denotes a nonperturbative
strong interaction scale that is finite as $N_c \rightarrow
\infty$.} and the wavefunction
is given to linear order in $r\lqcd$, while
the potential energy is given to quadratic order in
$r\lqcd$.  The spatial part of the wavefunction $\phi(\vec x)$
has a more rapid dependence on $r$ which will be computed later in this
article.
As $r$ goes from zero to infinity, $F(r)$ goes from
$-\pi$ to $0$.  Consequently, we expect that $F'(0)$
is positive and $F'''(0)$ is negative.  This is true for
example for the solution given in Ref.~\anw\
where a particular
four derivative term in the chiral Lagrangian for pion
self-interactions is used to stabilize the soliton.
Furthermore, the constituent quark model suggests that
$g$ is positive.  Thus, Eq.~\pot\ implies that the
$\Lambda_Q$ is bound by a harmonic oscillator potential
with $\kappa > 0$.

In the limit $N_c
\rightarrow \infty$, the nucleon is infinitely
heavy and terms in the Hamiltonian involving
the nucleon momentum are neglected.  The
lowest energy state then has the spatial
wavefunction $\phi(\vec x) = \delta^3(\vec x)$
corresponding to the minimum energy classical
configuration where the nucleon is located at
$\vec x = 0$.  The $1/N_c$ terms in the
Hamiltonian involving the nucleon momentum
give the spatial wavefunction a finite
spread and it is this finite extent which is responsible
for the Isgur-Wise function $\eta_0(v \cdot
v')$.  Terms of order $1/N_c$
involving the momentum are found by writing $\Sigma =
\Sigma(\vec x - \vec r(t))$ and quantizing the
collective coordinate $\vec r\,(t)$.  This procedure
yields
\eqn\hkin{
H_{\rm kin} ={{\vec p}^{\,2} \over {2
m_N}}-
{4\over 3} F'(0)\ \left(\vec I_H \cdot \vec I_N\right)\
{{\vec S_N \cdot \vec p} \over m_N},
}
where $m_N$ is the mass of the nucleon.
The first term is
the usual kinetic energy of the soliton and the
second term in Eq.~\hkin\ results from the second term
in Eq.~\hlag.  The Schr\"odinger equation is
\eqn\screq{
\left[ H_{\rm kin} + V_\Lambda(\vec x) \right] \Psi_{\Lambda}(\vec x)
= E \Psi_{\Lambda}(\vec x),
}
which implies that $\phi(\vec x)$ obeys the
differential equation
\eqn\difeq{
\left[ -{{\vec\nabla^2}\over {2 m_N}} + V_\Lambda(\vec x)
\right] \phi(\vec x) = E\ \phi(\vec x),
}
in the rest frame of the bound state.
Note that the term linear in $\vec p$ in Eq.~\hkin\ is necessary
for the wavefunction in Eq.~\lamstate\  to obey the
Schr\"odinger equation~\screq.  This result is not
surprising because both the term linear in
$\vec p$ and the part of the wavefunction linear in $\vec x$
arise from the peculiar definition of parity in Eq.~\hparity.
In deriving Eqs.~\screq\ and \difeq\ we have treated
the term linear in $\vec p$ in $H_{\rm kin}$
as a perturbation and
neglected its action on the ``small'' part of the
wavefunction $\Psi_{\Lambda}(\vec x)$ ({\it i.e.} the piece
proportional to $\vec x$). As we shall see shortly, for large $N_c$ the
term proportional to $\vec x$ in $\Psi_{\Lambda}(\vec x)$ and the term linear
in $\vec p$ in $H_{\rm kin}$ are subdominant and can be neglected in
the calculation of $\eta_0(v\cdot v')$.

In general, for large $N_c$ we expect the
potential $V_\Lambda(\vec x)$ to have the harmonic
oscillator form
\eqn\pothar{
V_\Lambda(\vec x) = V_0 + \frac 1 2 \kappa \vec x^2.
}
The absence of a term linear in $r$
requires $F''(0)=0$ which is a consequence
of the derivative expansion of the chiral
Lagrangian for pion self-interactions.  Quantum
corrections can induce nonanalytic behavior
in $\vec x$, but because of an explicit factor
of $1/ N_c$ such terms are less important than
those we have kept.  The particular expression
for $\kappa$ in Eq.~\kapdef\ is, however, model
dependent and arises from keeping only terms with
one derivative in the chiral Lagrangian for heavy
meson-pion interactions.

The model independence of the results of
this paper follows from
an analysis of large $N_c$ power counting.
In the limit $m_Q\rightarrow\infty$, the typical
size (or momentum) of
the bound state wavefunction occurs when
the kinetic energy ${\vec p}^{\,2}/2m_N$
and potential energy $\kappa{\vec x}^{\,2}/2$  of the bound state
contribute equally to the total energy $E-V_0$,
\eqn\typical{
r \sim \left(\kappa m_N\right)^{-1/4},\qquad
p \sim \left(\kappa m_N\right)^{1/4}.
}
Since $\kappa$ is of order $\lqcd^3$ and $m_N$ is of order $\lqcd N_c$,
the $N_c$ dependence of the typical binding energy is given by
\eqn\typenergy{
E-V_0 \sim \lqcd N_c^{-1/2}.
}
It is
now straightforward to see that higher order terms in the effective Lagrangian
can be neglected. Any term in the effective Lagrangian is a function of $\vec
r=\vec r_N-\vec r_H$, $\dot{\vec r}_N$, and $\dot{\vec r}_H$.
Terms involving only
the soliton field carry an overall factor of $N_c$.
These terms are independent of the relative
coordinate $\vec r$ and depend only on $\dot{\vec r}_N$. In the effective
Hamiltonian, the dependence of these terms on $\dot{\vec r}_N$
enters through powers
of the nucleon momentum.  A term with $n$ powers of the nucleon
momentum has the following large $N_c$ behavior,
\eqn\pterms{
\lqcd N_c \left({p\over M_N}\right)^n \sim \lqcd N_c N_c^{-3n/4}.
}
Thus, all terms with higher powers of the nucleon momentum than
the leading order kinetic energy term $p^2/ 2 m_N$ are suppressed
by more powers of $1/N_c$
and can be neglected in the large $N_c$ limit. Terms in the Lagrangian
which involve
the interaction between the soliton and the heavy meson can depend on $\vec r$,
but they are at most of order one in the large $N_c$ limit.
The typical scale of the $r$ dependent
interaction is $\lqcd$, so higher order interaction terms have the form
\eqn\rterms{
\lqcd \left(\lqcd r\right)^m \left({p\over M_N}\right)^n
\sim \lqcd N_c^{-m/4}N_c^{-3n/4}.
}
Hence, all the higher order interaction terms in the effective Lagrangian
other than the harmonic potential are higher order in $1/N_c$ and
can be neglected (including the term
linear in $\vec p$ in $H_{\rm kin}$ Eq.~\hkin).

There is also a class of $1/m_Q$ corrections which can be summed
exactly.
So far, we
have concentrated on the order of limits $m_Q\rightarrow\infty$
followed by $N_c
\rightarrow \infty$. We now switch to the situation in which
$m_Q,N_c\rightarrow\infty$ simultaneously, with the ratio
$\lqcd N_c/ m_Q$ held fixed.
The only additional term in the effective Lagrangian which must be included in
this double scaling limit is the kinetic energy of the heavy meson,
\eqn\extra{
{p^2\over m_H} \sim {m_N\over m_Q} {p^2\over m_N} \sim
\left({\lqcd N_c\over m_Q}\right)
{p^2\over m_N},
}
which is of the same order as a term we have included,
since $\lqcd N_c/ m_Q$ is of order one.
Higher order terms in $1/m_Q$ such as
\eqn\suchas{
{p^4\over m_Q^3} \sim \left({m_N\over m_Q}\right)^3 {p^4\over m_N^3},
}
are of order one times terms which can be neglected by the power
counting arguments of the previous paragraph, so they too can be
neglected. Additional $1/m_Q$ effects arise
from $1/m_Q$ operators in the heavy
quark effective theory. The
$P^*-P$ mass difference is of order $\lqcd^2/m_Q\sim
\left(N_c \lqcd/m_Q\right)\lqcd/N_c$ which is subleading in
$N_c$. Higher derivative operators in the current are also suppressed.
For example, the leading correction
\eqn\lamoper{
{1\over m_c}\Tr\
{\bar H}^{(c)}\, \larr{\Dsl}\,\gamma^\mu\left(1-\gamma_5\right)
\, H^{(b)}
\sim {\lqcd N_c^{1/4}\over m_c}
\sim \left({\lqcd N_c\over m_Q}\right) N_c^{-3/4},
}
since the typical momentum of the heavy quark in the baryon is of order
$\lqcd N_c^{1/4}$.  Thus, no additional terms other than the kinetic
energy of the heavy meson are relevant.

\newsec{$\lb\rightarrow\lc\,e^-\,\bar\nu_e$ Decay}

For non-relativistic $\lc$ recoil,
the matrix element of the weak current Eq.~\explicit\ in the $\lb$ rest frame
is
\eqn\evaluate{\eqalign{
&\bra{\lc(v',s')}\bar c\,\gamma^\mu \left(1-\gamma_5\right)b\ket{\lb(v,s)} =
\int d^3\pprime \int d^3\vec p\ \  \phi_c^*(\pprime\,)\,\phi_b(\vec p\,) \cr
&\quad\frac14\bra{N(-\pprime +m_N \vprime\,,s')}\left.
N(-\vec p\,,s)\right\rangle
\bra{D(\pprime + m_D \vprime\,)}c\,\gamma^\mu\left(1-\gamma_5\right)b
\ket{\bar B(\vec p\,)}+\ldots\cr
}}
where the ellipsis represents terms involving at least one vector
meson.
$\phi_c$ and $\phi_b$ are the
Fourier transform
of the ground state wave function
\eqn\phifou{
\phi_{c,b}(\vec p\,) = {1 \over {(\pi^2 \mu_{c,b}\, \kappa)^{3/8}}}
\exp\left({-\vec p^{\,2}/2\sqrt{\mu_{c,b}\, \kappa}}\right),
}
where $\kappa$ is defined in Eq.~\kapdef, and $\mu_{c,b}$ is the reduced mass
$\mu=m_N\,m_H/(m_N+m_H)$ of the bound state, with
$m_H=m_D,m_B$ for the $c$ and $b$ subscripts, respectively.
The $\Lambda_Q$ state is a superposition of products of $N$ with
$P$ and $P^*$ states. However, we do not need the details of the
Clebsch-Gordan structure\footnote{*}{The factor of $1/4$ is the square
of the appropriate
Clebsch-Gordan coefficient for the term displayed explicitly in Eq.~\evaluate.}
of the $\Lambda_Q$ state, as will become clear soon.

The nucleon matrix element vanishes unless
\eqn\pconst{
\pprime = \vec p + m_N \vprime,
}
at which point, for the term explicitly displayed in Eq.~\evaluate,
the required heavy meson matrix element is
\eqn\mesmat{
\bra{D\left(\vec p + (m_N+m_D)\vprime\,\right)}c\,
\gamma^\mu\left(1-\gamma_5\right)b
\ket{B\left(\vec p\,\right)}.
}
This matrix element can be evaluated in terms of heavy meson form factors.
In the $\Lambda_Q$ heavy meson-nucleon bound state,
the typical momentum is of order $\vev{p} \sim \left(
m_N \kappa \right)^{1/4}$ and so
form factors for semileptonic $\Lambda_b \rightarrow
\Lambda_c\, e^-\, \bar \nu_e $ decay are smooth functions of
$m_N\vprime/\left(m_N \kappa \right)^{1/4}$ which is of order $N_c^{3/4}
\vprime$. Consequently, we are interested in the kinematic region $v'$
of order $N_c^{-3/4}$, and so
for large $N_c$ we can replace the heavy meson form
factors by their rapidly varying part, $C_{cb}(v\cdot v')$, and neglect
the slow variation in the meson Isgur-Wise function.
This is why we do not need the
details of the Clebsch-Gordan structure of the $\Lambda_Q$ bound state.
The  $\Lambda_b \rightarrow
\Lambda_c\, e^-\, \bar \nu_e $ form factor in the large $N_c$ limit can be
written in terms of an Isgur-Wise function, as in Eq.~\explicit, with
\eqn\iweta{\eqalign{
\eta_0 &= \int d^3\vec p\ \ \phi^*_c(\vec p + m_N \vec
v^{\,\prime})\ \phi_b(\vec p\,) \cr
&= \left[{2 \left(\mu_c\,\mu_b\right)^{1/4}\over \sqrt \mu_b +
\sqrt \mu_c}\right]^{3/2}
\exp\left[- m_N^2 {\vec v}^{\,\prime\,
2}/2\sqrt\kappa\left(\sqrt\mu_b+\sqrt\mu_c\right)\right].
}}
Since for non-relativistic recoils
$\vec v^{\,\prime\,2} \approx 2(v\cdot v' - 1 )$, the
expression for $\eta_0$ in a general frame has the
form
\eqn\iwform{
\eta_0(v\cdot v')
= \left[{2 \left(\mu_c\,\mu_b\right)^{1/4}\over \sqrt \mu_b +
\sqrt \mu_c}\right]^{3/2}
\exp\left[- m_N^2 (v\cdot v'-1)/\sqrt\kappa\left(\sqrt\mu_b+\sqrt\mu_c
\right)\right].
}
The result we have obtained is valid in the heavy quark and
large $N_c$ limits, where we
have included all terms of order $(m_N/m_Q)^n\sim (N_c\lqcd/m_Q)^n$.
The baryon form factors for $\Lambda_b \rightarrow \Lambda_c
e^- \bar \nu_e$ are still written in terms of a single
function in this limit, even though we have included a class of $1/m_Q$
corrections. In the large $N_c$ limit, the function $\eta_0(v\cdot v')$
falls off rapidly away from zero recoil. Derivatives of $\eta_0$ at zero
recoil diverge as $N_c\rightarrow\infty$. Eq.~\iwform\ indicates that
the $m^{\rm th}$ derivative is of order $N_c^{3m/2}$, and
includes all contributions of this order neglecting less divergent pieces.
For example, there could be corrections to $\eta_0$ of the form
$N_c^{1/2} \left(v\cdot v'-1\right)$
in the exponent. This term is small compared with the leading term
$N_c^{3/2} \left(v\cdot v'-1\right)$, but is significant when $v\cdot v'-1$ is
of order $N_c^{-1/2}$.
Thus Eq.~\iwform\ is valid in the large $N_c$ limit in the
region where $v\cdot v'-1 \ltap \CO(N_c^{-3/2})$, \ie\ in the region where
$\eta_0$ falls from about unity to a very small quantity.

At zero recoil in the limit that
$m_b\rightarrow\infty$, $\eta_0$ has the form
\eqn\expandiw{
\eta_0(1) =
\left[{2 \left(\mu_c\, m_N\right)^{1/4}\over \sqrt m_N +
\sqrt \mu_c}\right]^{3/2} =1 - \frac3{64}\left({m_N\over m_D}\right)^2   +
\ldots,
}
where we have expanded the exact expression
in a power series in $m_N/m_D$. The term linear in $m_N/m_D$
vanishes, which is consistent with Luke's theorem \ref\luke{M.E. Luke,
\pl {252}{1990}{447}\semi
H. Georgi, B. Grinstein, and M.B. Wise, \pl{252}{1990}{456}\semi
C.G. Boyd and D.E. Brahm, \pl {257}{1991}{393}.}.
For physical values of $m_N/m_D$, the
correction to the symmetry limit prediction $\eta_0(1)=1$ is only 1\%.

The parameter $\kappa$
can be determined to be $(530\ \MeV)^3$
in the Skyrme model using the shape function used in
Ref.~\anw, and the value of $g$ obtained in Ref.~\glm. With this value of
$\kappa$, we find that the orbitally excited $\Lambda_Q$ state should be
about 400~MeV above the ground state, and that the form factor $\eta_0$
of Eq.~\iwform\ is
\eqn\iwnum{
\eta_0(v\cdot v')\sim 0.99\, \exp\left[-1.3 \left(v\cdot v'-1\right)\right],
}
using the known values of $m_N$, $m_D$ and $m_B$.
The Skyrme model prediction for $\kappa$ is sensitive to the precise
shape of the soliton solution. A better way to determine $\kappa$ is to
use the experimentally measured excitation energy of the orbitally
excited $\Lambda_Q$, which is $\sqrt{\kappa/\mu_Q}$.

The large $N_c$ predictions of this paper rely on the number of light
quarks in the heavy baryon being large. For $N_c=3$ there are only two
light quarks in $\Lambda_Q$, so we expect our results to only be
qualitatively correct. Nevertheless, it is interesting that the baryon
form factors are calculable in the large $N_c$ limit of QCD. There are
other results that can be computed using the methods developed here. For
example, the Isgur-Wise functions for transitions to excited states are
also calculable. It should also be possible to derive our results
using the methods of Witten \ref\witten{E. Witten, \np{160}{1979}{57}.}.
It would be interesting to explore that
approach.

\bigskip
\centerline{\bf Acknowledgements}
We would like to thank J. Hughes for several discussions which stimulated
our interest in the subject. E.J. and A.M. would like to thank the
Fermilab theory group for hospitality.
This work was supported in part by DOE grant \doe\
and contract \#DEAC-03-81ER40050,
and by a NSF Presidential Young Investigator award \pyiam.
\bigskip

\appendix{A}{The $\xi$ Basis}

In this appendix, we briefly discuss the computation of the binding potential
in the $\xi$ basis discussed in Ref.~\jmw. The notation is the same as found in
Ref.~\jmw.
The $\xi$ basis is singular at the origin, but has a simple transformation
rule for the $H$ field under parity,
$$
H(x^0,\vec x) \rightarrow \gamma^0 H(x^0,-\vec x) \gamma^0.
$$
The interaction Hamiltonian in this basis is
\eqn\hintxi{
H_I = -{ig\over 2}\int d^3 x\Tr \bar H H \gamma^j\gamma_5
\left(\xi^\dagger\partial_j
\xi- \xi\partial_j\xi^\dagger\right).
}
Expanding the Goldstone boson field
\eqn\xidef{
\xi_0 = \exp\left(i F(r)\ \hat x\cdot \vec\tau\right/2),
}
in a power series about $\vec x=0$, and using $\xi = A \,\xi_0\,
A^{-1}$, we get
the interaction potential
\eqn\potxi{\eqalign{
&\hat V_I(\vec x) = 2 g\ \Tr A \tau^j A^{-1} \tau^k\ \left\{
F'(0)\left(  S_{\ell H}\cdot \hat x\ I_H^k\ \hat x^j -{\frac 1 2}
 S_{\ell H}^j I_H^k \right)\right.\cr
&\ \left. + r^2 \left(\frac 1 {12} \left[F'(0)\right]^3-\frac 1 {12}
F'''(0)\right)
S_{\ell H}^j\ I_H^k + S_{\ell H}\cdot \vec x\
I_H^k\ x^j \left(-\frac 1 {12} \left[F'(0)\right]^3
+\frac 1 3 F'''(0)\right)\right\}.
}}
Note that only even powers of $x$ occur in Eq.~\potxi\ because the parity
transformation is trivial in the $\xi$ basis.
To leading order in $N_c$, the kinetic term of the soliton can be neglected, so
that states with a definite value of $A$ are eigenstates of the Hamiltonian. It
is convenient to consider the soliton state that is an eigenstate of $A$ with
$A=1$. States with other values of $A$ can be obtained by an isospin
transformation, and so have the same energy.
On states with $A=1$, the
interaction potential reduces to
\eqn\potaone{\eqalign{
&\hat V_I(\vec x) = 4 g\ \left\{
F'(0)\left(  S_{\ell H}\cdot \hat x\ I_H \cdot \hat x -{\frac 1 2}
 S_{\ell H}\cdot I_H\right) \right.\cr
&\ \left.+ r^2 \left(\frac 1 {12} \left[F'(0)\right]^3
-\frac 1 {12} F'''(0)\right)
S_{\ell H}\cdot I_H + S_{\ell H}\cdot \vec x\
I_H\cdot \vec x
\left(-\frac 1 {12}
\left[F'(0)\right]^3+\frac 1 3 F'''(0)\right)\right\}
}}
This Hamiltonian is singular at the  origin because of the coordinate
singularity in the
$\xi$ basis, and the eigenstates of the Hamiltonian will also be singular at
the origin. We know that the singularity at the origin has the form
$\vec\tau\cdot\hat x$, because that is the transformation function from the
singular $\xi$ basis to the non-singular basis used earlier in this article. We
therefore write the eigenstates of the interaction potential Eq.~\potaone,
$\ket{\psi}$ in terms of new eigenstates $\ket{\phi}$ which are related by the
unitary transformation
\eqn\statetrans{
\ket{\psi} = \left( 2\ \vec I_H\cdot \hat x \right) \ket{\phi}.
}
The interaction potential in the $\ket{\phi}$ basis is
\eqn\potphi{\eqalign{
\hat V_I'(\vec x) &= \left(2\ \vec I_H\cdot \hat x\right)\ \hat V_I(\vec x)\
\left(2\ \vec I_H\cdot \hat x\right)\cr
&=2g\ \left\{
F'(0)\ \vec S_{\ell H}\cdot \vec I_H + r^2 \left(-\frac 1 9
\left[F'(0)\right]^3 + \frac 5 {18}
F'''(0)\right)
\vec S_{\ell H}\cdot \vec I_H \right.\cr
&\left.\qquad + \left(\frac 1 6
\left[F'(0)\right]^3 + \frac 1 3 F'''(0)\right)
\left(\vec  S_{\ell H}\cdot \vec x\,
\vec I_H\cdot \vec x -\frac13 r^2
\vec S_{\ell H}\cdot \vec I_H \right)\right\},
}}
where we have used the relations
\eqn\relns{\eqalign{
&\left(\vec I_H \cdot \hat x\right)^2 = {\frac 1 4},\cr
&\left(\vec I_H \cdot \hat x\right) \left(\vec  S_{\ell H} \cdot\vec  I_H
\right)
\left(\vec I_H \cdot \hat x\right) =
{\frac 1 2} \left(\vec  S_{\ell H} \cdot \hat x\right)
\left(\vec I_H\cdot \hat x\right) - {\frac 1 4}
\left(\vec S_{\ell H} \cdot \vec I_H \right), \cr
}}
that follow from the anticommutation relation
\eqn\anticom{
I_H^j I_H^k + I_H^k I_H^j = \frac12\delta^{jk},
}
for the isospin-1/2 operators $I_H$. It is convenient to define the operator
$\vec K_H = \vec I_H + \vec S_{\ell H}$, in terms of which
Eq.~\potphi\ can be rewritten as
\eqn\potk{\eqalign{
\hat V_I'(\vec x)&=g\ \left\{
F'(0)\ \left(\vec K_H^{\,2}-\frac 3 2\right) + r^2 \left(-\frac 1 9
\left[F'(0)\right]^3 +\frac 5 {18}
F'''(0)\right)
\left(\vec  K_H^{\,2}-\frac 3 2\right) \right.\cr
&\left.\qquad\qquad+ \left(\frac 1 {6}
\left[F'(0)\right]^3 +\frac 1 3 F'''(0)\right)
\left(\vec K_H\cdot \vec x\
\vec K_H \cdot \vec x -{\frac 1 3}r^2\vec K_H^{\,2} \right)\right\}.
}}
The allowed values of $K_H$ for the $H$ field are $K_H=0$ and $K_H=1$ obtained
by
combining $I_H=1/2$ with $S_{\ell H}=1/2$. The states with $K_H=0$ are the
bound
physical baryon states
containing a heavy quark. On the $K_H=0$ states, Eq.~\potk\
reduces to the potential
\eqn\potzero{\eqalign{
\hat V_I^{K_H=0}(\vec x)&=-{\frac 3 2}g \left\{
F'(0) + r^2 \left(-\frac 1 9 \left[F'(0)\right]^3 +\frac 5 {18} F'''(0)
\right) \right\}\cr
&= -{\frac 3 2}g F'(0) + {\frac 1 2} \kappa r^2, \cr
}}
where $\kappa$ is defined in Eq.~\kapdef.
Thus the states with $K_H=0$ are bound in an attractive harmonic oscillator
potential, and the potential at the origin has the value $-3gF'(0)/2$. The
potential is more complicated for exotic states which have $K_H=1$. The last
term
in Eq.~\potk\ is an irreducible tensor operator with $K_H=2$ and contributes a
spin-orbit term to the interaction potential (but note that it does not cause
any mixing between $K_H=0$ and $K_H=1$ states). The spherically averaged
potential
for $K_H=1$ states is simple to compute,
\eqn\potone{\eqalign{
\hat V_I^{K_H=1}(\vec x)&=\frac 1 2 g\left\{
F'(0) + r^2 \left(-\frac 1 9 \left[F'(0)\right]^3 +\frac 5 {18} F'''(0)
\right) \right\}\cr
&= \frac 1 2 g F'(0) - \frac16 \kappa r^2 . \cr
}}
Thus the $K_H=1$ states are unbound since the potential at the origin is
positive, and the interaction potential is a repulsive inverted harmonic
oscillator potential.
The $\Lambda_Q$ state is obtained from the $K_H=0$ state by applying a
projection operator~\glm, so the interaction potential for the
$\Lambda_Q$ state is Eq.~\potzero. This is precisely the potential given
in Eq.~\pot\ of the text.

The kinetic term in the $\xi$ basis on the $A=1$ soliton states has the form
\eqn\kinxi{
\CL = {1\over 2} M_N \,\dot{\vec x}^2-{ 2\over r^2}\, \dot{\vec x} \cdot \left(
\vec x\times \vec I_H \right).
}
The first term is the usual soliton kinetic energy, and the second term is from
the expansion of
\eqn\laag{
\CL=
{i\over 2} \Tr \bar H H v^\mu \left( \xi^\dagger \partial_\mu\xi +
\xi\partial_\mu\xi^\dagger\right).
}
The kinetic Hamiltonian in the $\xi$ basis obtained from the Lagrangian
Eq.~\kinxi\ is
\eqn\kinxih{
H_{\rm kin} = {1\over 2M_N} \left(\vec p_N + {2\over r^2}\, \vec x\times
\vec I_H\right)^2.
}
Transforming from the singular basis using Eq.~\statetrans\ gives
the kinetic energy
\eqn\kinenergy{
H_{\rm kin}' = \left( 2\ \vec I_H\cdot \hat x \right)\ H_{\rm kin}
\ \left( 2\ \vec I_H\cdot \hat x \right) = {\vec p^{\,2}\over 2M_N}.
}
The bound state problem in the $\xi$ basis reduces to that of a
three-dimensional harmonic oscillator with a conventional kinetic term.

\listrefs
\bye